# Personalized Programming Education: Using Machine Learning to Boost Learning Performance Based on Students' Personality Traits

Chun-Hsiung Tseng[1], Hao-Chiang Koong Lin[2], Andrew Chih-Wei Huang[3], and Jia-Rou Lin[4]

## Abstract

Studies have indicated that personality is related to achievement, and several personality assessment models have been developed. However, most are either questionnaires or based on marker systems, which entails limitations. We proposed a physiological signal–based model, thereby ensuring the objectivity of the data and preventing unreliable responses. Thirty participants were recruited from the Department of Electrical Engineering of Yuan-Ze University in Taiwan. Wearable sensors were used to collect physiological signals as the participants watched and summarized a video; they then completed a personality questionnaire based on the big five factor markers system. The results were used to construct a personality prediction model, which revealed that galvanic skin response and heart rate variance were key factors predicting extroversion; heart rate variance also predicted agreeableness and conscientiousness. The results of this experiment can elucidate students' personality traits, which can help educators select the appropriate pedagogical methods.

## Motivation and Background

What is the significance of personality assessment models? Over the years, a variety of personality assessment models have emerged, including the big five model, self-efficacy and innovativeness, locus of control, and the need for achievement (Kerr et al., 2018). Many researchers have established a correlation between an individual's personality and their level of success. For instance, O'Connor and Paunonen emphasized the strong relationship between personality and academic performance (O'Connor & Paunonen, 2007). Meanwhile, Rothmann and Coetzer's study found a connection between job performance and personality traits among employees at a pharmaceutical company (Rothmann & Coetzer, 2003). In the field of adaptive learning, most research focuses on using parameters gathered from e-learning environments. However, as Brinton et al. pointed out, students' video-watching behaviors can provide insight into when they need help from instructors (Brinton et al., 2016). In traditional classroom settings, such information is typically unavailable, but students' personality traits may serve as a viable alternative.

[1] Department of Electrical Engineering, YuanZe University, R.O.C., lendle@saturn.yzu.edu.tw
[2] Department of Information and Learning Technology, National University of Tainan, R.O.C., koong@mail.nutn.edu.tw
[3] Deptartment of Psychology, Fo Guang University, R.O.C., chweihuang@mail.fgu.edu.tw
[4] Department of Electrical Engineering, YuanZe University, R.O.C., zoe841228@gmail.com

Assessment methods have already been developed for evaluating personality traits using established models. For example, Goldberg introduced a marker system that transparently quantifies one's degree of Extraversion, Agreeableness, Conscientiousness, Emotional Stability, and Imagination in terms of the Big Five model (Goldberg, 1992). Similarly, Sherer et al. devised a generalized scale for measuring self-efficacy (Sherer et al., 1982), and Craig et al. created a system to assess locus of control (Craig et al., 1984). However, these existing assessment tools have limitations and are mainly questionnaire or marker system-based. Despite their widespread use and validation, the need for additional assessment methods persists. As summarized by Akash Choudhury, using questionnaires has some limitations e.g. poor/unreliable/incomplete responses[5].

Our study proposes an approach based on physiological signals, which offers several advantages. Firstly, participants do not need to provide manual responses. Secondly, the use of physiological signals provides a more objective measure. Thirdly, as Tiwari et al. and Šalkevicius et al. have pointed out, people's physiological signals are linked to their psychological status, such as emotion and anxiety levels (Tiwari et al., 2019; Šalkevicius et al., 2019). Therefore, analyzing participants' physiological signal changes during events can potentially reveal valuable information about their internal psychological status and, by extension, their personality traits. This approach allows for a more in-depth understanding of individuals.

Furthermore, we conducted an experiment in Yuan Ze University R.O.C. to see whether or not delivering lecturing materials based on students' personality traits is helpful to enhance their learning performance. An elective course "Game Development", which has 22 students enrolled was chosen. Based on the research results, students who were given learning materials based on their personality traits outperformed other students in their learning results significantly.

## Related Works

Numerous models for assessing personality traits have been developed over time. Among the various personality models available, the big five model is particularly popular. Goldberg first proposed this model in 1990, and subsequently developed a marker system that transparently quantifies one's personality traits in terms of Extraversion, Agreeableness, Conscientiousness, Emotional Stability, and Imagination (Goldberg, 1992).

Rothmann and Coetzer discovered that there were positive associations between several personality traits and task performance (Rothmann & Coetzer, 2003). The healthcare sector in Pakistan conducted a survey and found that Agreeableness, Conscientiousness, and Openness to Experience positively impacted organizational effectiveness (Butt et al., 2020). O'Connor and Paunonen established that Conscientiousness was strongly linked to academic success (O'Connor & Paunonen, 2007). Duff et al. found that individuals with different personality traits tend to choose different learning strategies (Duff et al., 2004).

Previous research has demonstrated that physiological signals can be utilized as indicators of psychological states. Egger, Ley, and Hanke conducted a survey that found physiological signals to be 79.3% accurate in assessing individuals' emotional states, while speech recognition achieved 80.46% accuracy in detecting happiness and sadness specifically (Egger et al., 2019). Sriramprakash et al. successfully developed a model to assess an individual's stress level using electrocardiogram and GSR(Sriramprakash et al., 2017). Wache conducted a study to examine the relationship between personality traits and different physiological signals while participants watched emotional movie clips (Wache, 2014). In Bastos' research, multiple models were created to evaluate personality traits through physiological signals such as pupil, ECG, BVP, and EDA (GSR). A total of 473 features were extracted from these signals. The findings indicated that EDA and BVP were the best predictors for Openness, ECG and EDA for Agreeableness, and ECG and EDA for Extraversion (Bastos, 2019). The research of Butt et al. extracted 11 features from participants' EEG, GSR, and PPG (Photoplethysmography) signals for assessing their personality traits and the classification accuracy was from 67% to 92% (Butt et al., 2020).

---

[5] https://www.yourarticlelibrary.com/social-research/data-collection/questionnaire-method-of-data-collection-advantages-and-disadvantages/64512

# The Development of the Personality Assessment Model

## Experiment Design

The study consisted of two phases. In the first phase, physiological signals were collected from participants to construct a personality assessment model. In the second phase, participants were provided with learning materials tailored to their personality traits to evaluate if this information could improve their learning performance. Thirty participants were recruited from the electrical engineering department at YuanZe University in Taiwan for the first phase. Participants were selected randomly and were not informed about their personality traits, grades, or interests prior to the study. To collect participants' personality traits and physiological signals, an experimental protocol was designed as follows:

1. attach GSR and heart rate sensors to participants; during the experiment, these sensors will collect participants' physiological signals and saved them to a local disk
2. participants watched a video of 8 minutes and 30 seconds which is about JavaScript programming
3. after watching the video, participants were asked to summarize the content of the video
4. participants completed the big five personality traits models questionnaire

Then, we invited students enrolled the "Game Development" to join the second phase experiment. Participated students were then divided into the control and the experiment groups. For students in the experiment group, we deliver learning materials based on their personality traits while for students in the control group, their learning materials were randomly delivered. In the end of the semester, the learning performance of the two groups were compared.

## Personality Trait Collection

Participants' personality traits were collected using the IPIP big five factor markers system developed by Goldberg[6]. The system employs a 50-item questionnaire, with each question offering five response options from very inaccurate to very accurate. It assesses the tendency scores of the big five personality traits. Participants with higher scores in a specific personality trait are considered to have a stronger tendency towards it. To enhance convenience, a web application was created to administer the questionnaire, and a screenshot of the web application is provided below:

---

[6] https://ipip.ori.org/new_ipip-50-item-scale.htm

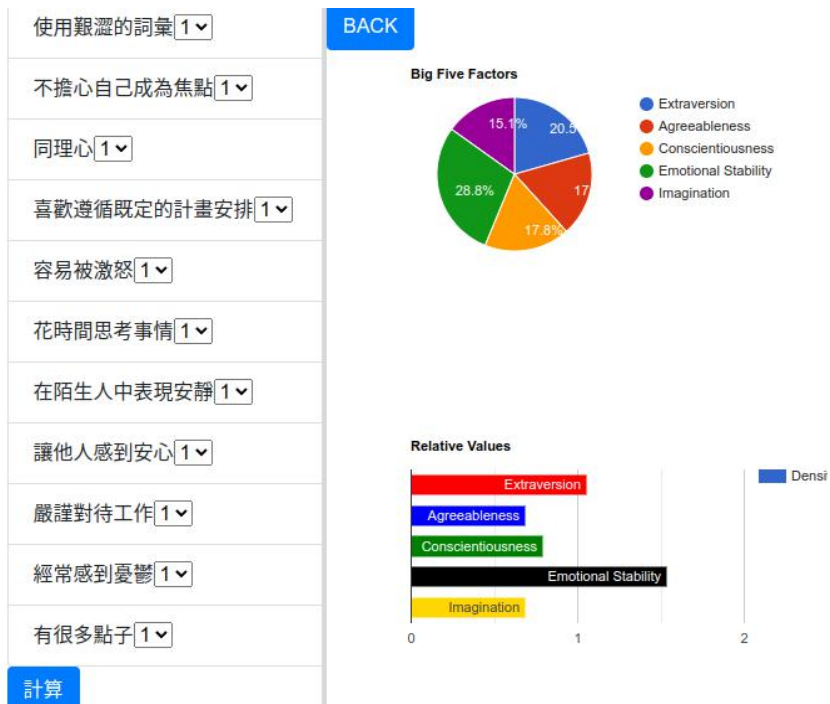

Figure 1 The screenshot of the Web application

In addition to hosting the questionnaire, the Web application also shows the level of the tendency of each personality trait compared to other participants.

Physiological Signals Collection

The grove GSR sensor[7] and the grove ear clip heart rates sensor[8] were installed on Raspberry Pi for the collection of physiological signals. The figure below shows the device and the usage during the collection step:

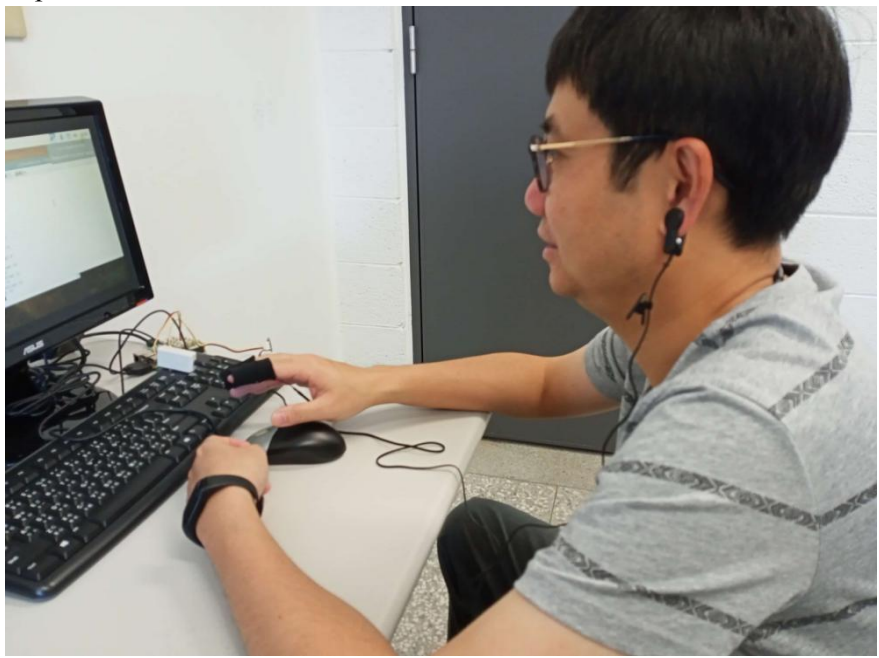

---

[7] https://wiki.seeedstudio.com/Grove-GSR_Sensor/

[8] https://wiki.seeedstudio.com/Grove-Ear-clip_Heart_Rate_Sensor/

Figure 2 The usage scenario of the device

We also built a GUI-based application for the collection of physiological signals. The application was written using the JavaFX technology and can be run on the Raspberry Pi board. The figure below shows the application:

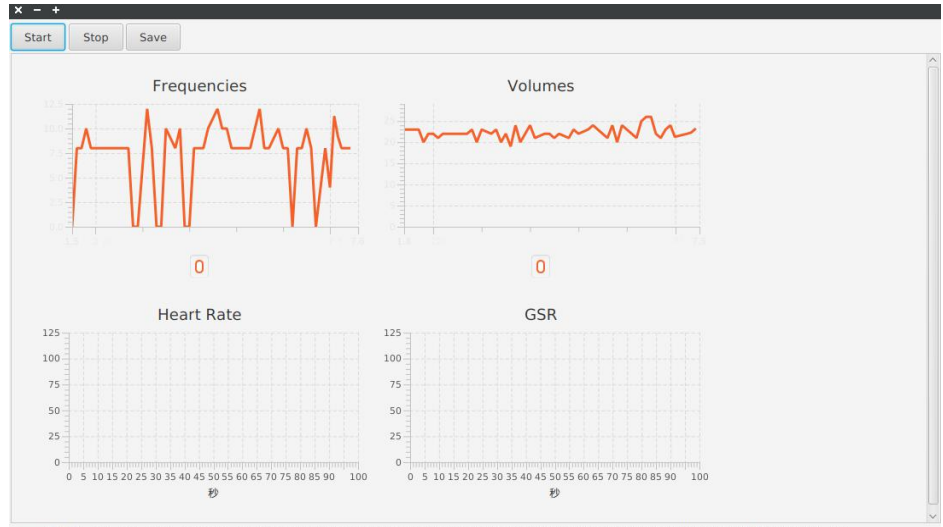

Figure 3 The physiological collection application

The application collects voice frequencies, volumes, heart rates, and GSR values. However, in this research, only the heart rate values and GSR values were kept.

## Data Analysis

### Preprocessing of Data

Due to the different processing speeds of the sensor modules, different sampling rates are used:
- GSR values: 0.3 samples per second
- heart rate values: 1.1 samples per second

This is a test. For each participant, we recorded the physiological signals for the first 7 minutes and equally divided the data into three segments. To reduce noise, data below the 10th percentile and data higher than the 90th percentile were removed. Then, for each segment, we calculated the variance of each signal. Additionally, we calculated the difference between the variance of GSR values in different segments and labeled them as delta_gsr12, delta_gsr23, and delte_gsr13, respectively. The same procedure was also applied to the variance of heart rate values, and they were labeled as delta_hr12, delta_hr23, and delta_hr13, respectively. The variance values were normalized using the z-score method and the delta values were represented as ratio values, that is, delta_gsr12 was calculated as

{variance of GSR in the 2nd segment}-{variance of GSR in the 1st segment} / {variance of GSR in the 1st segment}

Participants' personality trait values were preprocessed using the steps below:
- the median values of each personality trait were calculated

- the personality trait values were then labeled as "H" and "L" according to whether they were higher than the corresponding median value or not

## Statistics of Raw Data

First, the figures below show the distribution of GSR and heart rate variance in each segment:

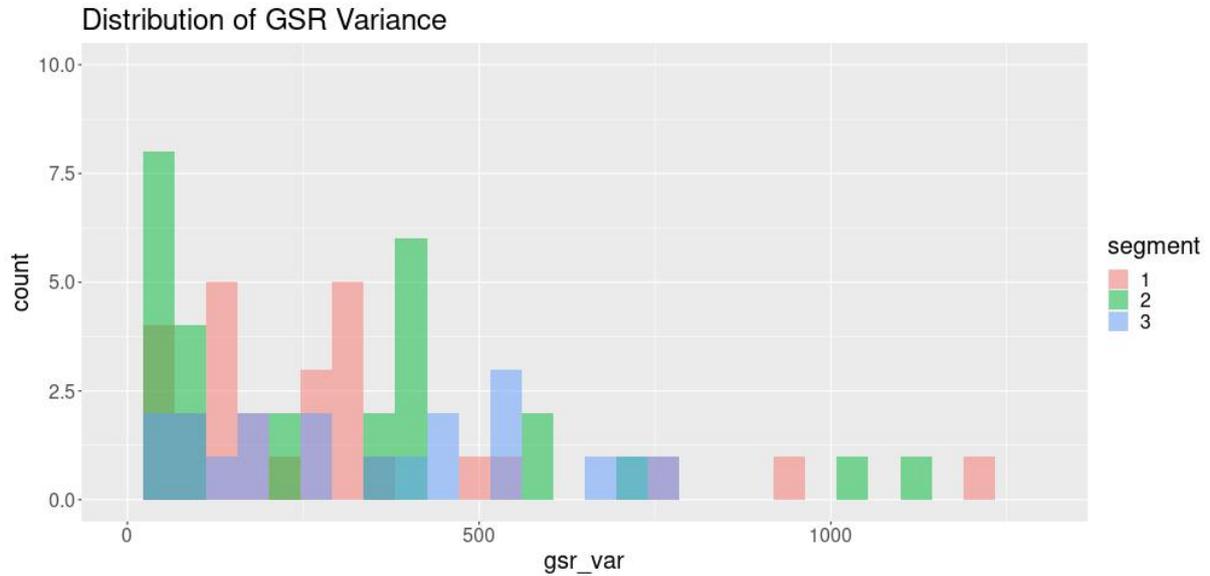

Figure 4 Distrubution of GSR variance

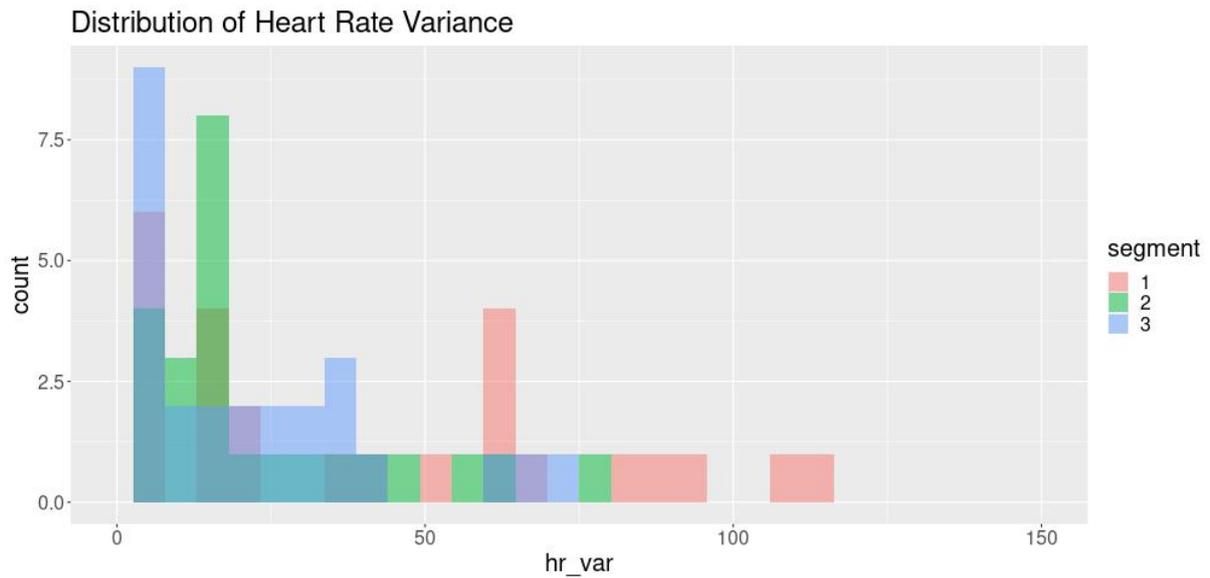

Figure 5 Distribution of heart rate variance

The table below lists the descriptive statistics of the input values (not normalized):

Table 1 The un-normalized input value

|  | mean | sd | median | min | max | skew |
|---|---|---|---|---|---|---|
|  |  |  |  |  |  |  |
| gsr1_var | 284.66 | 278.54 | 244.87 | 4.64 | 1204.09 | 1.69 |
| hr1_var | 55.95 | 104.59 | 20.85 | 0.73 | 566.88 | 3.98 |
| gsr2_var | 281.39 | 300.72 | 203.63 | 3.37 | 1102.54 | 1.21 |
| hr2_var | 18.76 | 20.33 | 13.49 | 0.52 | 77.44 | 1.32 |
| gsr3_var | 325.27 | 520.75 | 176.74 | 2.55 | 2723.01 | 3.32 |
| hr3_var | 20.52 | 20.82 | 8.64 | 0.74 | 70.56 | 1.05 |
| delta_gsr12 | 0.45 | 1.98 | 0.13 | -0.97 | 8.48 | 2.64 |
| delta_gsr23 | 0.39 | 1.47 | -0.16 | -0.94 | 4.69 | 1.45 |
| delta_gsr13 | 0.93 | 4.13 | -0.20 | -0.98 | 21.44 | 4.21 |
| delta_hr12 | -0.16 | 0.77 | -0.47 | -0.98 | 1.54 | 0.93 |
| delta_hr23 | 1.04 | 3.47 | 0.10 | -0.90 | 18.17 | 4.11 |
| delta_hr13 | 0.11 | 1.39 | -0.26 | -0.94 | 4.58 | 2.09 |

Then, the descriptive statistics of the input values (normalized) are listed in the table below:

Table 2 The normalized input value

|  | mean | sd | median | min | max | skew |
|---|---|---|---|---|---|---|
| gsr1_var | 0.00 | 1.00 | -0.14 | -1.01 | 3.30 | 1.69 |
| hr1_var | 0.00 | 1.00 | -0.34 | -0.53 | 4.89 | 3.98 |
| gsr2_var | 0.00 | 1.00 | -0.26 | -0.92 | 2.73 | 1.21 |
| hr2_var | 0.00 | 1.00 | -0.26 | -0.90 | 2.89 | 1.32 |
| gsr3_var | 0.00 | 1.00 | -0.29 | -0.62 | 4.60 | 3.32 |
| hr3_var | 0.00 | 1.00 | -0.57 | -0.95 | 2.40 | 1.05 |
| delta_gsr12 | -4.07 | 19.42 | -0.10 | -103.64 | 7.64 | -4.63 |
| delta_gsr23 | -0.39 | 1.35 | -0.33 | -3.91 | 3.49 | 0.25 |
| delta_gsr13 | 0.32 | 5.25 | -0.39 | -10.02 | 19.48 | 1.58 |
| delta_hr12 | -2.43 | 11.66 | 0.38 | -61.21 | 5.66 | -4.38 |
| delta_hr23 | -0.21 | 2.26 | 0.01 | -6.56 | 6.61 | -0.01 |
| delta_hr13 | -1.96 | 7.25 | -0.33 | -21.52 | 16.57 | -0.53 |

The figure below illustrates the distribution of each personality traits:

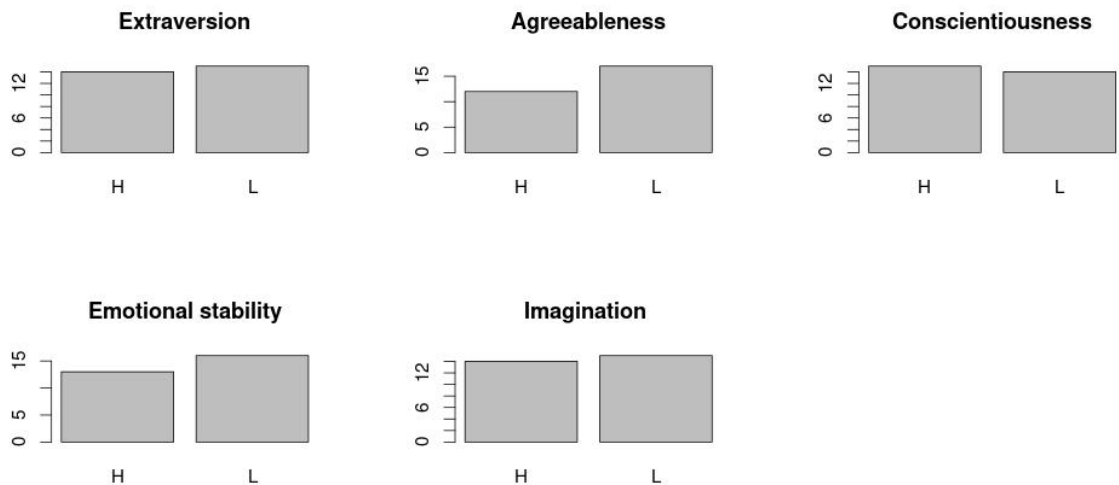

Figure 6 The distribution of personality traits

## The Prediction Models

Since the range of GSR values and heart rate values are different, we used the scaled dataset for the construction of the prediction model. The input/output variables are listed below:
- input variables: gsr1_var, hr1_var, gsr2_var, hr2_var, gsr3_var, hr3_var, delta_gsr12, delta_gsr23, delta_gsr13, delta_hr12, delta_hr23, delta_hr13
- output variables: Ex_class, Co_class, Es_class, Ag_class, Op_class

The goal of the prediction model is to use the input variables to predict the level of the output variables. The Random Forest[9] algorithm was used for constructing the model. The R implementation[10] of the Random Forest algorithm was used. To achieve better prediction precision, we had to figure out the best set of input variables for model construction. The R package Boruta[11] was used for input variable selection, while tuneRF[12] was used for tuning the parameters for the Random Forest implementation. Each model was trained separately, and the results are described below.

Ex_class

To predict Ex_class, we first used the Boruta package to figure out the important variables. The list below shows the output of the execution of Boruta:

*Boruta performed 38 iterations in 0.8774686 secs.*
 *3 attributes confirmed important: delta_gsr12, delta_hr12, gsr2_var;*
 *9 attributes confirmed unimportant: delta_gsr13, delta_gsr23,*
*delta_hr13, delta_hr23, gsr1_var and 4 more;*

As suggested by Boruta, delta_gsr12, delta_hr12, gsr2_var were important variables to predict Ex_class. In the resulting Random Forest model, the mean decrease accuracy value of the three variables were 50.73191, 46.04577, and 48.88196, respectively, which showed that removing these variables will decrease the resulting accuracy greatly. The resulting OOB (Out-Of-Bag) error was 10.34%. The figure below shows the first tree in the resulting Random Forest model:

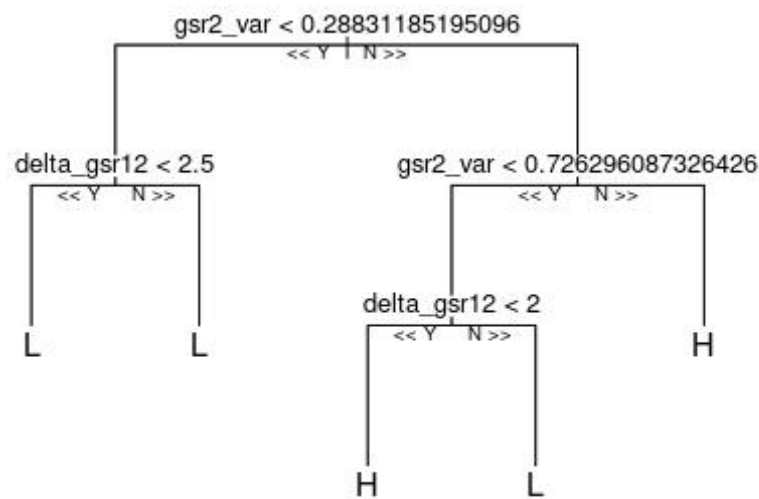

Figure 7 The random forest for Ex_class

---

[9] https://en.wikipedia.org/wiki/Random_forest
[10] https://www.rdocumentation.org/packages/randomForest/versions/4.7-1/topics/randomForest
[11] https://cran.r-project.org/web/packages/Boruta/Boruta.pdf
[12] https://www.rdocumentation.org/packages/iRF/versions/2.0.0/topics/tuneRF

Co_class

To predict Co_class, Boruta suggested that important variables were delta_gsr23, delta_hr13, and gsr2_var, and the corresponding mean decrease accuracy were 12.84836, 30.57563, and 28.50746, respectively. The resulting OOB error was 24.14% and the figure below shows the first tree of the resulting model.

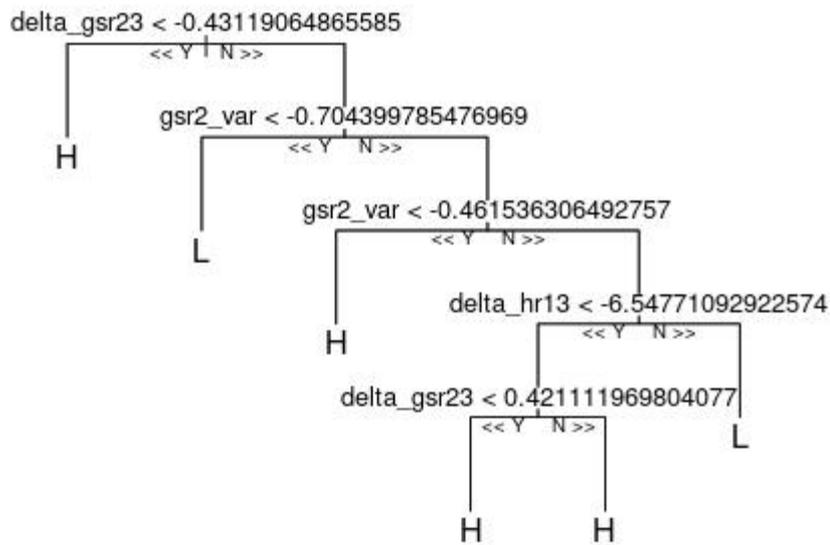

Figure 8  The random forest for Co_class

Ag_class

To predict Ag_class, Boruta suggested that important variables were Ex_class, Co_class, and hr2_var, and the corresponding mean decrease accuracy was 23.55775, 37.66948, and 14.26147, respectively. The resulting OOB error was 27.59% and the figure below shows the first tree of the resulting model.

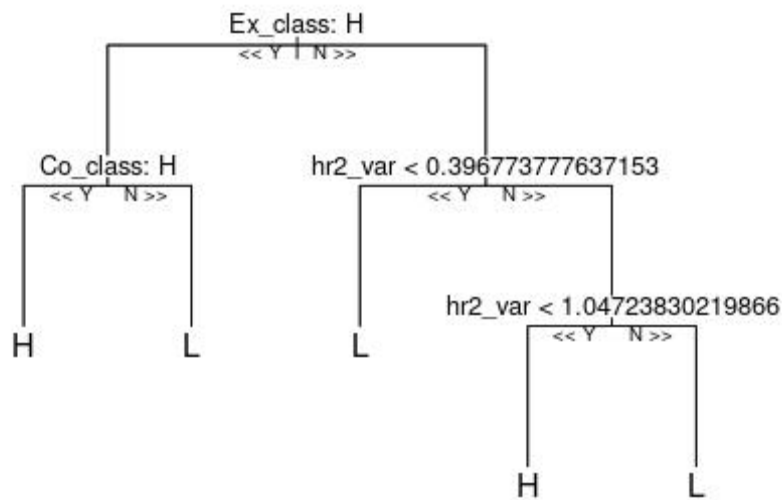

Figure 9  The random forest for Ag_class

### Es_class and Op_class

For Es_class and Op_class, we did not find any effective model in the experiment. However, among the 30 participants, we did find some tendencies. For example, we found that most participants with low Ex_class, high Co_class, and low Ag_class also had low Es_class. The table below summarizes the results:

Table 3 The observation of Es_class and Op_class

| Known Classes | | | Observed Tendency | | | |
|---|---|---|---|---|---|---|
| Ex_class | Co_class | Ag_class | Es_class | | Op_class | |
| | | | H | L | H | L |
| L | L | L | 50% | 50% | 25% | 75% |
| L | H | L | 75% | 25% | 75% | 25% |
| L | H | H | 67% | 33% | 67% | 33% |
| H | L | L | 25% | 75% | 25% | 75% |
| H | L | H | 100% | 0% | 100% | 0% |
| H | H | H | 29% | 71% | 57% | 43% |

## Enhance Learning Performance using Personality Information

We conducted an experiment in Yuan Ze University R.O.C. to see whether or not delivering lecturing materials based on students' personality traits is helpful to enhance their learning performance. An elective course "Game Development", which has 22 students enrolled was chosen. The students were then randomly distributed into the experiment group and the control group. After the midterm exam, 6 students withdrawn and did not finish the experiment. In the end, there were totally 10 students in the experiment group and 6 students in the control group. During the semester, we followed the Duff's research results which concluded that students tend to choose learning strategies (deep learning, surface learning, and strategic learning) based on their personality traits (Duff et al., 2004). We prepared 4 labs which students had to complete in 6 weekrs with learning materials designed based on the three learning strategies. Students in the experiment group received learning materials based on their personality traits while students in the control group received random learning materials. We conducted a before test and an after test to know students' understandings of the JavaScript language, and then calculated the delta of points they gained in the two tests to evaluate their improvements. The equation below were used:

$$\text{delta} = \frac{A_t - B_t}{B_t}$$

In which At and Bt refers to the points they obtained in the after and before test, respectively. The result is shown in the table below:

Table 4 The descriptive statistics of the two groups

|  | Group | N | Mean | Median | SD | SE |
|---|---|---|---|---|---|---|
| delta | Experiment | 10 | 0.560 | 0.600 | 0.123 | 0.0387 |
|  | Control | 6 | 0.356 | 0.400 | 0.153 | 0.0625 |

Then, a normality test using the Shapiro-Wilk method was conducted, the result is shown below:

Table 5 The normality test

|  | W | p |
|---|---|---|
| delta | 0.777 | 0.001 |

Note. A low p-value suggests a violation of the assumption of normality

Based on the result of the normality test, the data did not distribute normally. Then, both Student's t test and Mann-Whitney U test were performed:

Table 6 Student's t test and Mann-Whitney U test

|  |  | Statistic | df | p | Mean difference | SE difference | 95% Confidence Interval Lower | 95% Confidence Interval Upper |
|---|---|---|---|---|---|---|---|---|
| delta | Student's t | 2.95 | 14.0 | 0.011 | 0.204 | 0.0693 | 0.0558 | 0.353 |
|  | Mann-Whitney U | 8.00 |  | 0.016 | 0.200 |  | 0.1000 | 0.433 |

The non-parametric Mann-Whitney U test was used due to the non-normal distribution of the data, and results indicated a significant difference between the two groups ($p < 0.05$), with the experiment group scoring significantly higher than the control group. Despite the small sample size of 16 participants, the results suggest that the difference is significant.

# Conclusions and Future Work

In this research, we proposed a mathematical model to assess participants' personality traits via their physiological signals. Using the proposed mechanism is more convenient than the original personality traits assessment method, which is questionnaire-based. Knowing one's personality traits has some advantages. For example, the connection between one's personality traits and task performance has already been established in the research of Rothmann and Coetzer (Rothmann & Coetzer, 2003).

Our findings are partly consistent with those of Bastos (2019), who found that ECG and EDA can be used to evaluate an individual's degree of extraversion. In our study, we discovered that an individual's GSR and heart rate variance are the most significant factors in predicting their extraversion level. This similarity can be attributed to the fact that ECG is highly correlated with heart rate variance (as mentioned in https://www.ncbi.nlm.nih.gov/pmc/articles/PMC7472094/), and EDA is equivalent to GSR (as explained in https://en.wikipedia.org/wiki/Electrodermal_activity). Moreover, like Bastos, we found that ECG and EDA can be employed to assess an individual's degree of agreeableness. Our results also highlight the significance of heart rate variance in assessing an individual's agreeableness level.

Compared with Wache's results, both Wache's model (Wache, 2014) and the proposed model showed that heart rate variance is related with one's degree of agreeableness and conscientiousness. Wache's model showed the correlation between one's EEG and degree of emotional stability, but we did not collect EEG signals in our experiment. Besides, Wache's model found that one's degree of creativity (openness to experience) was best predicted by GSR, while our model did not show such a relationship. Please note that the research of Wache focused more on finding the correlation between physiological signals and degree of personality traits, while the proposed model focused more on building a prediction model.

In addition to building a prediction model and a software system, An experiment was conducted to examine whether the adoption of the proposed method enhances the academic performance of students. According to our experimental findings, the utilization of educational materials tailored to align with the personality traits of students resulted in improvements in their academic performance. Due to the limited number of participants in the experiment, the results may not be widely generalizable, but they still hold some degree of reference value.

## Acknowledgment

This research is partially supported by the "Development and implementation of algorithms for detecting learning difficulties related to personality traits: Taking university programming courses as the research subject." project, which was funded by the National Science and Technology Council, Taiwan, R.O.C. under Grant no. 111-2410-H-155-002.

# Conflict of interest

The authors declare that they have no known competing financial interests or personal relationships that could have appeared to influence the work reported in this paper.

# Data availability

Data sharing not applicable to this article as no datasets were generated or analyzed during the current study.

# Ethical approval

The research was approved by National Cheng Kung University Governance Framework for Human Research Ethics in Taiwan. The case number was 110-174 and the research results passed the final examination on 2022/7/28. These data are accessible online via the url: https://rec.chass.ncku.edu.tw/application/search?combine=110-174. The state is "close" now, which shows the research was performed in accordance with relevant guidelines enforced by the institute. The examination process adhered to the rules showed here: https://rec.chass.ncku.edu.tw/en.

# Informed consent

Informed consent was obtained from all participants and/or their legal guardians. We provided a consent document which participants have to sign before participating in the research.